# Prospects for hydrogen storage in graphene

Valentina Tozzini[*] and Vittorio Pellegrini

*NEST-Istituto Nanoscienze – Cnr and Scuola Normale Superiore,*
*Piazza San Silvestro 12, 56127 Pisa Italy.*

**Abstract**

Hydrogen-based fuel cells are promising solutions for the efficient and clean delivery of electricity. Since hydrogen is an energy carrier, a key step for the development of a reliable hydrogen-based technology requires solving the issue of storage and transport of hydrogen. Several proposals based on the design of advanced materials such as metal hydrides and carbon structures have been made to overcome the limitations of the conventional solution of compressing or liquefying hydrogen in tanks. Nevertheless none of these systems are currently offering the required performances in terms of hydrogen storage capacity and control of adsorption/desorption processes. Therefore the problem of hydrogen storage remains so far unsolved and it continues to represent a significant bottleneck to the advancement and proliferation of fuel cell and hydrogen technologies. Recently, however, several studies on graphene, the one-atom-thick membrane of carbon atoms packed in a honeycomb lattice, have highlighted the potentialities of this material for hydrogen storage and raise new hopes for the development of an efficient solid-state hydrogen storage device. Here we review on-going efforts and studies on functionalized and nanostructured graphene for hydrogen storage and suggest possible developments for efficient storage/release of hydrogen at ambient conditions.

## I. Introduction

Hydrogen is currently considered one of the most promising "green" fuel, owing to the fact that its staggering energy content of 142 MJ $kg^{-1}$ exceeds that of petroleum by a factor of three and that the product of its combustion is water vapour. However, it should be remembered that hydrogen is not an energy source, but a secondary energy carrier. It means that hydrogen must be produced, and exactly the same amount of energy needed in the production process is subsequently released during its use in fuel cells. Consequently the advantage of hydrogen for energy must be carefully considered with respect to other carriers, such as electricity.

In light of this, the issue of finding ways and materials for efficient hydrogen storage assumes a primary importance. Compared to electricity, in fact, hydrogen can potentially solve the problem of energy dispersion, because once energy is chemically stored then, in principle, it can be indefinitely conserved and transported with no dispersion. Practically the problem of energy dispersion is not eliminated, but transformed in a problem of matter (hydrogen) efficient confinement.

During the last decades several means for hydrogen storage were considered. It is important to recall that the efficiency of storage is usually measured by two parameters: the gravimetric density, GD, namely the weight percentage of hydrogen stored of the total weight of the system (hydrogen+container), and the volumetric density, VD, that is the stored hydrogen mass per unit volume of the system. Both parameters are important, since for practical application a hydrogen storage device must be both light and compact. Thus the possible storage systems can be evaluated plotting their GD and VD in a Cartesian diagram, where the good ones occupy the upper right corner. This is done in Fig.1. Even the current Department of Energy-USA (DoE) targets are expressed in terms of VD and GD (green crosses in Fig.1): specifically, for on-board hydrogen storage systems for light-duty vehicles, the expected hydrogen gravimetric capacity should reach in 2015 the level of 5.5% and volumetric capacity 0.04 $Kg/m^3$, which would correspond to an usable energy per mass of 1.8kWh/$Kg^1$.

The most obvious storage system, i.e. the compression of $H_2$ in tanks, has the advantage of having maximum GD (100%). Disadvantages of this storage method are that it needs low temperatures and/or high pressures, and therefore it raises issues related to safety. In addition, the VD of liquid hydrogen is not the maximum possible. In fact, it corresponds to the density of liquid hydrogen in molecular form of 0.068 kg/l, while hydrogen can be further compacted within solid state systems in the form of hydrides. The storage of atomic hydrogen was indeed realized in several solid systems including transition metal hydrides and light metal hydrides, which display VD between 0.08 and 0.15 $Kg/m^3$ (see Figure 1). The two main drawbacks of these systems are their relatively high weight (low GD) and the fact that they need extreme temperature and pressure conditions in order to control hydrogen adsorption/desorption[2,3]. Recently several studies have focussed on $MgH_2$, which displays good hydrogen uptake-release kinetic especially if doped with transition metals[4].

Another class of recently considered compounds are the hydrocarbons and N- B- hydrides[5,6,7] that certainly satisfy the GD and VD requirements, and require chemical reaction to control hydrogen charge/discharge. Graphene, the two-dimensional one-atom-thick crystal composed by carbon atoms arranged in honeycomb geometry, when considered as a medium for hydrogen storage, falls in this last class, and consequently it inherits some of the advantages. The $sp^2$ covalent-bonding arrangement of the carbon atoms in the honeycomb structure allows efficient binding to hydrogen atoms[8]. In addition graphene is stable and robust and, therefore, can be easily transported for long distances. At the same time it is mechanically flexible enabling new charging/discharging strategies at room conditions that exploit the dependence of hydrogen-carbon binding on local curvature[9]. In perspectives, graphene's flexibility and unique electronic properties could allow integration of a hydrogen-storage module into flexible and light, all-graphene based devices.

---
[*] Corresponding author: Fax: +39 050 509 417; Tel: +30 050 509 43; E-mail: v.tozzini@sns.it

In this paper we illustrate the state of-the-art of graphene system for hydrogen storage, and highlight new potentialities and possible future developments.

## II. Hydrogen storage with graphene: state of the art

### I.1 Physisorption versus Chemisorption on single graphene layers

Hydrogen can basically be adsorbed on graphene in two different ways: either by physisorption, i.e. interacting by Van Der Waals (VdW) forces, or by chemisorption, i.e. by forming a chemical bond with the C atoms.

Physisorption usually happens with hydrogen in molecular form. The $H_2$ binding energy was theoretically evaluated in the range 0.01-0.06 eV[9,10], this large spread of values depending on the fact that London dispersion forces are very elusive and difficult to represent. Despite that, it is clear that the binding of molecular hydrogen is very weak and requires therefore low temperatures and high pressures to ensure reasonable storage stability. It was shown by a simple empirical argument that in the most favourable conditions (high pressure and low temperature) $H_2$ can form a uniform compact monolayer on the graphene sheet, corresponding to GD of 3.3%[11] (doubled if the two sides are considered). The VD indeed depends on the possibility of compacting graphene sheets in complex structures, which is discussed in the next section.

Chemisorption of molecular hydrogen on graphene presents rather high barriers of approximately ~1.5 eV[12], because it requires the dissociation of $H_2$ (dissociative adsorption). Conversely, the chemisorption of atomic hydrogen is a rather favourable process: indeed commonly accepted values for H binding energy and chemisorption barriers are ~0.7eV and ~0.3eV, respectively. These values were extracted from several experimental studies on highly oriented graphite[13,11], and from theoretical studies mainly based on Density Functional Theory (DFT) with model systems mimicking graphite [14,15,16,17] or graphene[18]. More recent scanning tunnelling microscopy (STM) experiments have focussed on atomic-scale imaging of adsorption and clustering phenomena of hydrogen atoms on graphite[19,20,21,22]. The theoretical studies have shown, in particular, that adsorption of the first H atom locally modifies the graphene structure favouring further H binding, with a collective stabilization effect[23,24]. The formation of "dimers" of H on the graphene surface was shown to bring up to ~1eV gain in energy with respect to isolated bound H[25,26]. Atomic hydrogen absorption on epitaxial graphene on SiC was also investigated by STM showing formation of dimer structures, preferential adsorption of protruding graphene areas and clustering at large hydrogen coverage[27,28].

From the point of view of hydrogen storage, the maximum GD reachable in graphene with chemisorption is 8.3% (=1/12), i.e. even larger than the "ultimate" goal of DOE (see Fig 1). This corresponds to the formation of a completely saturated graphene sheet, with 1:1 C vs H stoichiometry, namely "graphane", whose stability was first hypothesized in a DFT-based theoretical study[29], and subsequently studied in experiments[8]. The experimental work, in particular, showed one-side hydrogenation of a graphene sheet and its reversibility by thermal annealing. Graphane is a strong insulator and removal of hydrogen in selected regions can lead to a controllable band-gap opening[30,31,32]. As in the case of physisorption, the VD depends on the possibility of building compact structures with graphene (or graphane) sheets.

The energy profiles for the processes of adsorption of hydrogen on graphene are summarized in Fig. 2.

### I.2 Hydrogen storage in nano-structured graphene

Being a single graphene layer a quasi 2D system, its VD is not well defined, thus in the evaluation of the potentialities of graphene for hydrogen storage one should consider graphene multi-layers, three-dimensional assemblies or nano-structures of graphene. As an example, an interesting prediction regarding the physisorption capabilities of a two-layer graphene system was derived from calculations based on hybrid post-Hartree-Fock/empirical potentials and including quantum treatment for hydrogen[10]. It was shown that both the GD and VD depend on the inter-layer separation, with highest values obtained for an interlayer separation of 6-8 Å (the graphite interlayer distance is 3.4 Å). In these conditions, the physisorption energy is nearly doubled with respect to the monolayer case, reaching values of ~0.1eV, because the attractive VdW forces of two layers combine together, inducing a sort of nanopump effect. This is capable of increasing the internal pressure of hydrogen with respect to the external one, reaching a high level of compression. Correspondingly, the GD is increased by ~30-40% of its single layer value, potentially reaching 8% at high pressure and low temperature, but remaining in the range of 3-4% at room temperature and high pressure. The model was subsequently improved by using more accurate interaction potentials and by exploring a wider range of temperature and pressure[33]. The optimal interlayer distance for hydrogen physisorption within the two layers (or equivalently, average diameter for nanoporous systems) was confirmed to be ~6Å leading to an "optimal" average density of the storage system roughly one half of that of graphite (see Fig 1, orange shaded strip). These additional theoretical studies also confirmed that the value of GD of 3% could be reached at room temperature only with high pressures, while at low temperature much higher values can be obtained, even exceeding the compression typical of liquid $H_2$. From the experimental side it was shown that such a layered structure can be realised by using graphene oxide and the interaction between hydroxyl groups and boronic acids[34]. This resulted in a sequence of graphene oxide layers connected by benzenediboronic acid pillars displaying an inter-layer separation of 10 Å with a predicted gravimetric capacity of around 6% at 77K at a pressure of 1 bar. It should be stressed that the natural interlayer distance in graphene oxide of 7 Å is not large enough for molecular hydrogen uptake owing to the presence of the epoxies and hydroxyl groups.

VdW interaction self-enhancement can be similarly postulated in any hollow graphene nano-structure. For example, an empirical estimate of the maximal VD vs GD relation of hydrogen physisorbed in nanotubes can be obtained assuming a level of compression similar to that of liquid $H_2$ and a full occupation of the cavity, and it is represented Fig. 1 (orange line). Simulations performed both with

DFT based methods and empirical force field substantially agree with this line, locating the GD of single walled nanotubes of ~1nm and ~2 nm at ~4% and 12% respectively[35,11]. These values are substantially confirmed by a number of experiments, but correspond to the optimal values reached at high pressures (~10Mpa) and low temperatures (~80K). The room conditions values are much lower. In fact, carbon nanotubes have been extensively considered as hydrogen storage media after the initial report of a GD of 5%[36]. Today, however, the best reproducible results yield a GD best value of around 1% at a pressure of 120bar at room temperature[37].

A similar hydrogen-storage capability was theoretically associated to an artificial three-dimensional structure composed by graphene layers placed at an interlayer distance of 12 Å and stabilized by carbon nanotubes inserted perpendicularly to the graphene planes[38]. *Ab initio* and Monte Carlo simulations have shown the stability of such pillared graphene structures, demonstrating that GD vs VD relation lies approximately on the same line as nanotubes. The simulated structures, in particular, display a GD of 8% at low temperature and high pressure, which decreases by an order of magnitude at room conditions, but can be recovered up to 6% at room temperature and ambient pressure after doping the pillared structure with Li cations. On the experimental side, molecular hydrogen adsorption in graphene-like nano-sheets obtained by chemically reducing exfoliated graphite oxide has been studied leading to an molecular hydrogen adsorption capacity of 1.2% at 77K and a pressure of 10bar and 0.68% and ambient pressure[39], extending previous controversial experiments on graphitic nanofibers[40]. A GD value of 2.7% at 25bar and room temperature was reported in graphene sheets obtained from graphene oxide after ultrasonic exfoliation in liquid[41].

I.3 Improving hydrogen storage by chemical functionalization

Though physisorption within layered or nanostructured graphene can potentially lead to reasonable storage GD and GV, it appears that large storage capacity values meeting the DoE goals are reached in very un-practical environmental conditions. In order to enhance the hydrogen adsorption at ambient conditions, therefore, various approaches including chemical functionalization of graphene were theoretically proposed. Most of the theoretical effort has been devoted to investigate the role of chemical functionalization in enhancing the molecular hydrogen adsorption at room temperature and ambient pressure conditions. In light of the weak interaction between molecular hydrogen and graphite/graphene, several routes were theoretically investigated to enhance both the binding and the gravimetric/volumetric storage capacity. One approach exploits the chemical decoration of graphene with alkali atoms such as Li, Na and K. In the case of Li it was shown that each adsorbed Li on graphene and nanostructured graphene can adsorb up to four $H_2$ molecules amounting to a gravimetric density above 10 wt%[42,43]. A change in hydrogen binding energy on K-decorated graphene was also predicted[44]. By means of DFT calculations, additionally, it was demonstrated that the combined effect of N-doping of graphene and application of an electric field normal to the sheet could induce dissociative adsorption of $H_2$[45]. A different approach is based on the exploitation of the (σ-π/d) Kubas interaction[46] to bind the hydrogen on active metal sites on the graphene surface. The advantages for hydrogen storage of this particular chemical interaction, half a way from chemisorption and physisorption, are related to optimization of hydrogen-graphene interaction for room-temperature application. Particularly studied are the cases of chemical functionalization of graphene with different transition metals such as Sc, V and Ti[47,48,49]. Several first-principle and DFT simulations showed the capacity of Ti atoms, for example, to bind up to four hydrogen molecules yielding, if attached to both graphene sides, a nominal gravimetric capacity of above 7%[47,48]. It was also predicted that stable ethylene-titanium complex should be able to attach up to ten hydrogen molecules yielding a further increase of hydrogen storage capacity[50]. An interesting experimental result was obtained by decorating graphene obtained by reduction from graphene oxide with Pd atoms. An increase of the GD at 30 bar at room temperature from 0.6% to 2.5% was reported and associated to a spillover mechanism in which hydrogen molecules undergoes dissociative chemisorption on the Pd atoms followed by H migration on the graphene layer[51]. The "spillover" effect, therefore, provides a mechanism for hydrogen molecule dissociation into hydrogen atoms. As we shall discuss in the next session this is relevant since atomic hydrogen adsorption/desorption on graphene can be finely controlled offering new strategies for hydrogen storage.

Transition metal elements, however, display large cohesive energies and then prefer to cluster[52], en effect detrimental to the hydrogen storage capacity although strategies to avoid clustering effects, for instance by applying strain to the graphene layer, have been proposed[53]. For this reason, decoration with Ca atoms has been also considered[54]. In particular recent theoretical studies have proposed Ca-decorated graphene nanoribbons for hydrogen adsorption reaching gravimetric capacities of 5% with negligible clustering of Ca atoms[55]. Other approaches to favour the anchoring of transition metals on graphene substrates have been explored. Among them recent theoretical simulations have suggested the potentialities of graphene oxide as a substrate where transition metals such Ti atoms attach to epoxies and hydroxyl groups offering a promising route to gravimetric capacities of around 5%[56].

## III. "Mechanically controlled" hydrogen storage on graphene

The following issues emerge from the previous sections: (i) compared to metal hydrides and other media for hydrogen storage, graphene-based systems have the advantage of being relatively light having good mechanical properties and high surface areas. GD and VD are not excellent but sufficient for the DoE goals. (ii) Within graphene-based systems, the advantages of hydrogen chemisorption with respect to physisorption relay on a much higher stability of the adsorbates, which makes the system very suitable for long time storage or transportation. In addition, at variance with molecular-based media (e.g. hydrocarbons or N and B based compounds) this is a real "solid state" system with advantages related to robustness and compactness with respect to liquids and gas phase systems.

There are, however, drawbacks linked to the chemical nature of the interaction (which are, in fact, in common with chemical based storage systems). Both processes of chemisorption and desorption of molecular hydrogen on pristine flat graphene have energetic barriers of the order of eV. Thus any efficient hydrogen storage device must consider ways to overcome, lower, or bypass such energy barriers.

**III.1 Graphene stretching and buckling: effect on H binding**

The extended nature of a graphene layer and its large flexibility suggest the possibility of lowering the energy barriers to hydrogen adsorption by means of structural manipulations. We recall that the effect of strain on adsorption of atomic[57] and molecular[58] hydrogen was recently studied by DFT methods. Both works have indicated that a strain up to ~10-15% tends to stabilize the adsorbate up to values of ~1eV per atom. The impact of strain on the chemisorption/desorption/hopping barrier is less clear. Compression producing an in-plane deformation of the sheet, however, seems to have both the effects of stabilizing the adsorbate and lower the barrier for associative desorption (or dissociative adsorption).

Considering the out of plane deformations, the recent experimental observation that hydrogen clusters are preferentially located on protruding areas of graphene on SiC[27,28] is in agreement with the previous observation that atomic H binding energy is increased (and barrier decreased) on fullerenes and nanotubes[59]. The empirical explanation of this effect is simple: convex surfaces have the $sp^2$ system distorted towards $sp^3$, which makes the protruding $\pi$ orbitals more reactive and prone to bind H. The accurate quantification of this effect, however, turns out to be quite complex, involving the interplay between mechanical distortion and change of the electronic structure due to curvature[60]. Indeed the effect of curvature of graphitic nanostructures and in particular carbon nanotubes has been investigated for a long time[61]. The chemisorption on the surface of nanotubes and fullerenes was studied in a number DFT-based theoretical papers in the last decade[62,63,64,65,66,67,68,69,70]. All together these works lead to the conclusion that the binding energy of hydrogen on the external surface of nanotubes and fullerene is increased considerably compared to the flat surface (graphene) by values of the order of 1-2 eV/atom (depending on the tube diameter and length, on the chirality, on the coverage and type of decoration). Additionally, in small nanotubes (less than (5,5)) the binding energy is such that even the binding of molecular hydrogen (i.e. via dissociative adsorption) is exothermic. Results about the barriers for adsorption are less clear: the barrier for the adsorption of atomic hydrogen is very small already on flat graphene and disappears for small convex curvature, while the barrier for dissociative adsorption of $H_2$, thought decreasing with curvature, remains in the range of eV. Thus the process can occur only in specific conditions. For instance, under high pressure the nanotube can undergo distortion creating more reactive sites. The enhancement of chemical reactivity of a graphene sheet with artificial hemispherical ripples of different radii was subsequently investigated again with DFT[71]. The ripples (unstable in the absence of hydrogen), were shown to be stabilized by H adsorption, whose binding energy was found to increase for larger corrugation leading to a hydrogen adsorbate stable with respect to free molecular hydrogen for extreme values of the corrugation.

The above results were rationalized by us by a systematic DFT study of the binding energy of hydrogen as a function of curvature, evaluated on intrinsically rippled graphene structures, stabilized by lateral compression, and displaying a wide range of curvatures[9]. We reported a linear increase of the binding energy of atomic H after choosing the value of the local curvature as the independent variable, instead of the average curvature of the ripple (see Fig.3). The binding energy values for fullerenes and nanotubes lie on the same line. Finally the binding energy of molecular hydrogen can be estimated to follow a similar behaviour, further enhanced by an amount dictated by the cooperative effect of the hydrogen already bound.

We found that also physisorption of molecular hydrogen depends on the curvature, but in an opposite way: VdW interactions are stronger in the concavities, due to the superposition of attractive basins of different adjacent carbon atoms. However, this effect is detectable only at low temperatures, since the related energies are just fraction of eV.

**III.2 Tuning hydrogen storage with changes of graphene curvature**

The results discussed above raises the possibility of exploit graphene's curvature for implementing practical ways to store/release hydrogen. Similarly the use the ripples generated by lateral compression was recently proposed to engineer graphane-graphene heterostructures[72] displaying finely tunable electronic properties.

Concerning hydrogen storage, one of the key problems to be solved for practical applications is related to the realization of a chemisorption/desorption mechanism that works at room conditions, without the recourse to extreme temperatures and pressures. Specifically, while it is relatively easy to chemisorb at least atomic hydrogen, the desorption process requires overcoming the associative barrier that is of the order of eV. To this end we have recently proposed to exploit the control of curvature of graphene to desorb hydrogen[9] The process is schematically depicted in Fig 4: After corrugation of graphene (e.g. by lateral compression), H atoms are loaded (1. of Fig 4) and preferentially bind on convexities (2. of Fig 4). In this configuration hydrogen on graphene can be safely transported, with minimal or no dispersion, being the C-H binding very stable. When requested, hydrogen can be forced to desorb by simply inverting the curvature of graphene, i.e. by transforming the convexities in concavities. In this way H atoms will find themselves in the concave part of the corrugated graphene (3. of Fig 4). Since in this configuration C-H binding is unstable, hydrogen will spontaneously desorb (back to 1.). We have explored this process in a time-dependent Car-Parrinello molecular dynamics simulation in which the curvature inversion is realised by the passage of a transverse acoustic phonon. Hydrogen is seen to desorb and form molecules, the system being able to get around the associative desorption barrier like a sort of "mechanical catalysis" effect (Fig 5, a).

These findings that combine the large flexibility of graphene layers with large and tuneable C-H biding energy, suggest a microscopic working scheme that sets the basis for an hydrogen storage/release device operating at fixed temperature and pressure that exploits the chemisorption as the storage mechanism and can reach estimated gravimetric densities of 8%.

# IV. Towards a practical device

The ideal hydrogen storage device should be cheap and robust, store large quantities of hydrogen as set by the DoE[1], have a short recharge time, and operate at room conditions. We have briefly reviewed in Section I the different options available so far. Concerning solid-state systems, in particular, current on-board metal hydride storage devices can store up to 7-8% hydrogen by weight but are costly, require a relatively long recharge time (10–20 min) and operate under large changes in temperature. An advantage is their relative compactness and the relatively low charging pressure as compared with compressed gas cylinders), which at this time represents a practical solution. Graphene, on the other hand, is an attractive medium for hydrogen storage because it is readily available also in large quantities and potentially at low cost. We have seen in Section II that its high surface area combined with the increased adsorption sites after chemical functionalization leads to high storage capacities establishing graphene as a promising candidate material for hydrogen storage.

A number of approaches based on hydrogen storage in various types of graphene structures and nanostructures are currently studied mostly on theoretical grounds. For those that possess sufficiently large storage capacity, heat must be applied to release hydrogen. In the case of applications on vehicles, this would involve the use of an on-board burner and heat exchanger.

In an attempt to avoid the requirements of changes in temperatures and pressures, we have proposed in Section III an approach in which hydrogen storage and release can be obtained by means of the control of graphene's curvature. While our simulations suggest the feasibility of this idea and the potentialities for realistic hydrogen-storage applications, there are still a number of practical issues to be solved.

In the loading phase our simulations show selective chemisorption of convex surfaces for atomic hydrogen. The energetic data (see Figure 4) demonstrate that also chemisorption of molecular hydrogen can be exothermic for very large local curvatures. However, in general, at variance with the case of atomic hydrogen, the chemisorption barrier is likely to remain quite high. In light of this, therefore, if hydrogen is not previously cracked (a process that requires a well-defined amount of energy), additional catalytic mechanisms are necessary such as, for example, the spillover mechanism produced by functionalization with Pd[51]. Besides chemical catalysts that would require functionalization of graphene often detrimental to its intrinsic properties, one could consider using an external electric field. Indeed, recent DFT-based calculation have shown that the associative desorption barrier can be decreased or even eliminated by applying high values of an electric field orthogonal to the graphene sheet[45]. Additionally, such electric fields can create or stabilize graphene's curvature[73]. We note that the combined effect of an electric field and an external compression leading to graphene rippling has not been investigated so far, but it is likely to produce cooperative effects and possibly to allow spontaneous adhesion even of molecular hydrogen.

On the experimental side the demonstration and full characterization of the predicted selective chemisorption of atomic hydrogen on convex parts of a corrugated graphene layer is in progress[74]. These experiments are currently carried out on graphene buffer layer grown on SiC that displays a natural corrugation due to interaction with the substrate[75] although a factor of ten less than that explored theoretically. In these experiments the loading of atomic hydrogen on graphene is monitored by atomically-resolved imaging with a scanning tunnelling microscopy in a ultra-high vacuum environment.

Concerning the desorption phase, the simulation results presented above suggest a possible desorption mechanism that requires the control of the curvature. To this respect, several options can be considered. In our simulation the curvature inversion is obtained by a transverse wave, which could be induced by coupling the system to a piezoelectric substrate. A different possibility is to functionalize the system with optically active organic molecules such as functionalized azo dyes whose size is controlled by an external radiation at specific wavelength. These would serve also as spacers for the multilayer system, solving at the same time the problem of increasing the volumetric capacity of the system (see Fig 5, (b)).

Also one should keep in mind that the curvature enhances the chemical reactivity of graphene in general, particularly to oxygen. Thus the loading phase must be realized in a pure hydrogen atmosphere, to avoid loading of other substances. Still in the loading phase, if atomic hydrogen is used, its pressure must be tuned in order to avoid competition with the recombination in molecules. In the desorption phase, on the other hand, the mechanical distortion of the system might produce dissipation of heat that must be kept at minimum. All these practical issues do not pose fundamental problems to the working principle of the devices but must be carefully considered for the optimization of the device.

## V. Conclusions

This review summarizes the state of the art concerning the use of graphene and graphene-based structures for hydrogen storage. From the available studies, graphene emerges as a promising material for hydrogen storage in terms of GD and VD densities. The possibility of creating functionalized graphene nanostructures and its peculiar characteristics such as large electrical conduction, robustness, manageability and flexibility open interesting scenario for its exploitation in future hydrogen technology. Particularly promising for hydrogen storage applications is the fact that graphene can be now produced on large and cost-effective scale by either top-down (such as exfoliation from bulk) or bottom up (atom by atom growth) techniques. Indeed large graphene oxide or graphene flakes can be produced from the exfoliation of pristine graphite. For example graphene flakes up to several micrometers in sizes can be obtained at high concentration (few mg/ml) by liquid phase exfoliation[76,77,78,79] and higher concentration (>5mg/ml) can be achieved in ionic liquid[80].

In this perspective paper we have also reported and quantified the tunability of the hydrogen binding energy as a function of graphene local curvature. The large variation of H binding energy makes the chemisorption a favourable process on convex sites and hydrogen release a favourable process on graphene concave sites. On the basis of these predictions we have suggested a multilayer graphene storage system that might lead to a hydrogen device that exploits the controlled change of the local curvature for storing and releasing hydrogen and can operate at room conditions with fast kinetics.

## Aknowledgements

We thank Ranieri Bizzari, Francesco Bonaccorso, Camilla Coletti, Sarah Goler, Stefan Heun, Pasqualantonio Pingue, Vincenzo Piazza and Marco Polini, for useful discussions and suggestions. CINECA supercomputing center resources were obtained by means of INFM-Progetto di Calcolo Parallelo 2009 and Platform "Computation" of IIT (Italian Institute of Technology). Partial support from MIUR through the program "FIRB - Futuro in Ricerca 2010", is also acknowledged.

## Figures

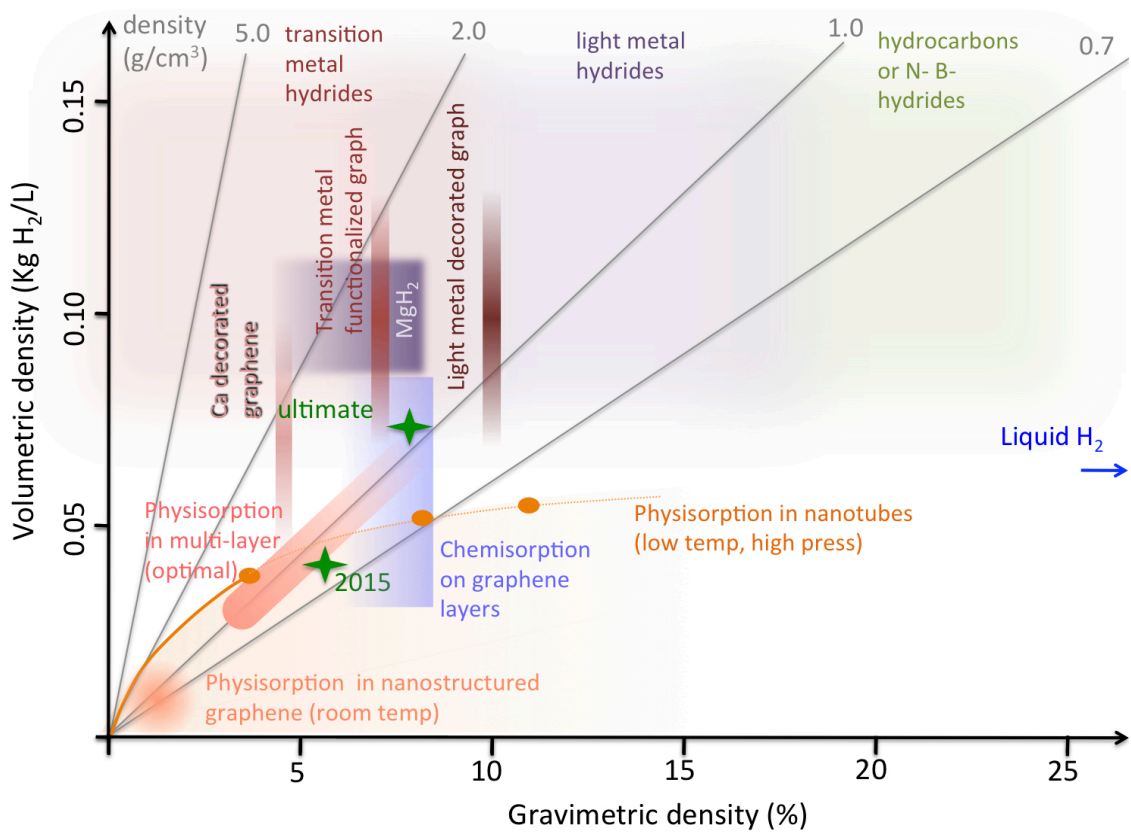

**Figure 1**
Gravimetric vs volumetric density diagram for several hydrogen storage systems including the graphene-based ones. The orange line represents the optimal relationship for physisorption in nanotubes (the dots correspond to different sizes). The line tends to the value of liquid hydrogen for large sized nanotubes. In general nanostructured physisorption based graphitic systems occupy the area below this line. The oblique shaded strip represents the optimal physisorption within graphene multilayers with spacing nearly double than in graphite (and density nearly one half). Different storage densities in this case correspond to different pressure and temperatures. The vertical dark red strips represent adsorption in decorated or functionalized graphene. These systems have been mostly studied at the level of a single layer, for this reason only the gravimetric density is well defined, while the volumetric density range has been roughly estimated considering variable inter-layer spacing 2-4 times that of graphite. The same criterion has been used to estimate the volumetric density for chemisorption in multilayers (blue rectangle); for this system the gravimetric density has a sharp right edge, corresponding to the maximum loading with 1:1 stoichiometry of C and H (~8%). The storage properties of systems based on material different from graphene (different metal hydrides (including $MgH_2$), hydrocarbons, N- and B- hydrides) are also reported as shaded areas in red, violet and green. The DOE targets (for 2015 and ultimate) are indicated with green stars. The constant density lines are in grey.

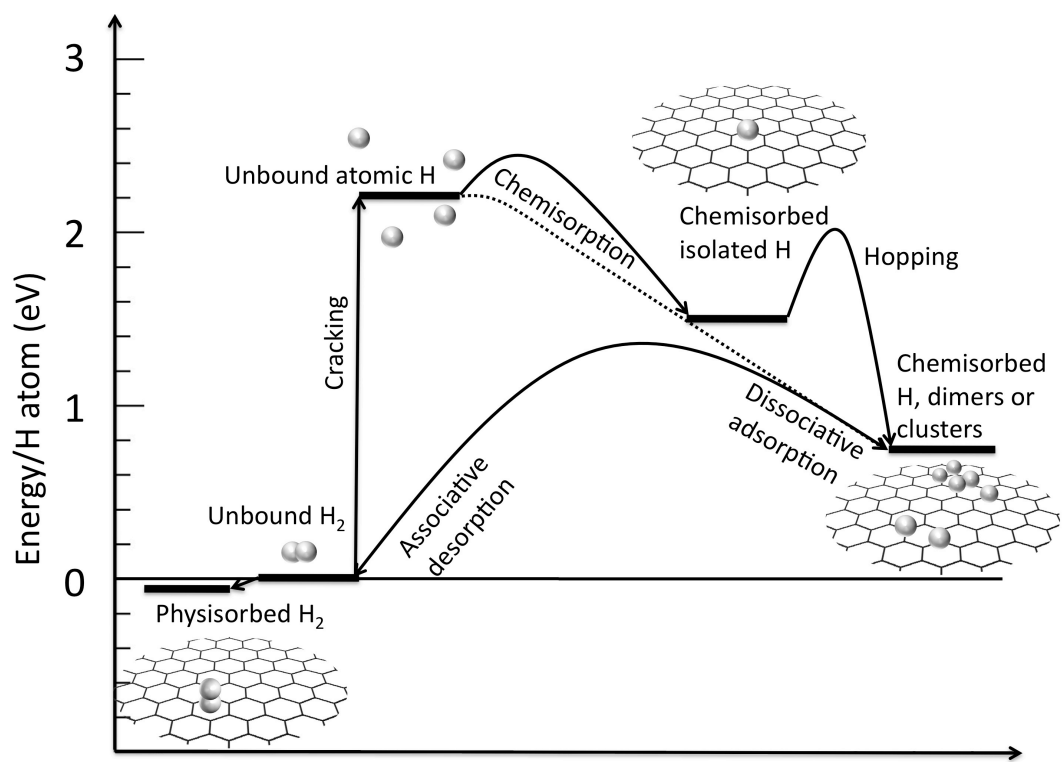

**Figure 2**
Energy level diagram for the graphene-hydrogen system. The energy is in eV *per H atom*, i.e. to obtain the values per $H_2$ each energy and barrier value must be doubled. Values of energy levels and barriers are deducted both from experimental and theoretical evaluations with average values taken when different values are available. The reference level is the pristine graphene plus unbound molecular hydrogen.

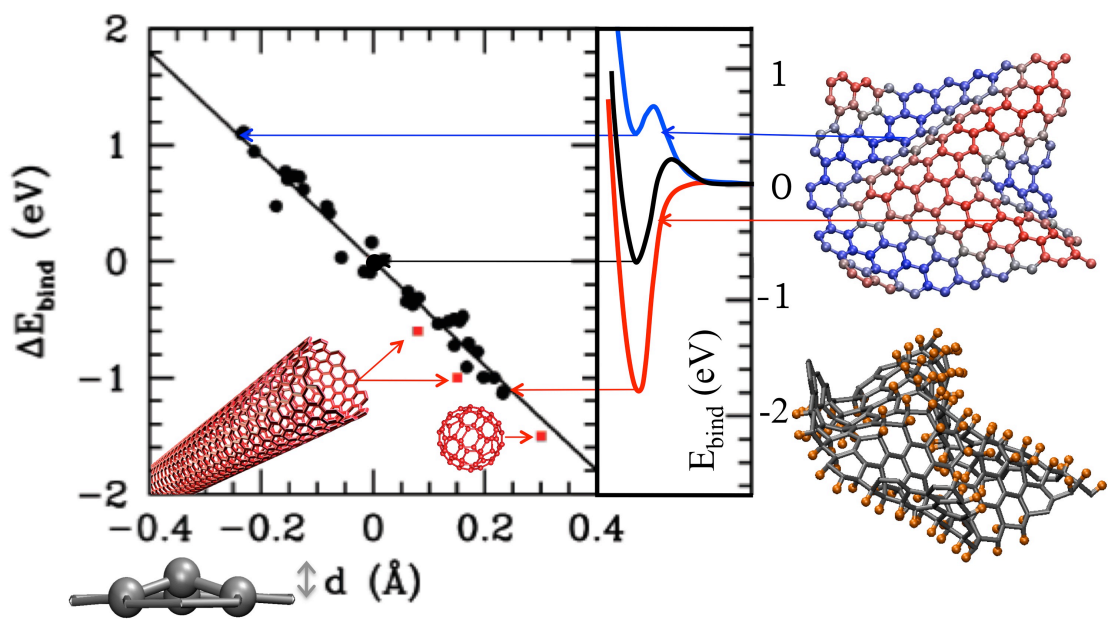

**Figure 3**
Calculated C-H binding energy versus the local curvature. The local curvature is measured by the puckering distance d of a given C atom with respect to the plane formed by its three first neighbours before H binding (see ball&sticks representation under the X axis). The variation of binding energy with respect to the flat graphene is reported on the left Y axis. The black dots correspond to the binding energy of isolated H evaluated on sampled C sites of the corrugated graphene sheet, shown on the right, with the convex parts in red and the concave ones in blue. The fitting black line is $\Delta E_{bind}[eV]= -4.449\, d[\text{Å}]$. Binding energy profiles are shown for three sample site on the convex (red), flat (blak), and concave (blue) regions respectively, as indicated by the arrows on the right (in this case the Y axis report the absolute binding energy using as a reference system the graphene plus isolated detached atomic H). The arrows on the left indicates the corresponding dots in the main graph. The red squares correspond to literature data of the binding energy on $C_{60}$ and on nanotubes of different lengths. The structure of graphene sheet hydrogenated on convexities is also reported in balls&sticks (gray = C, orange = H).

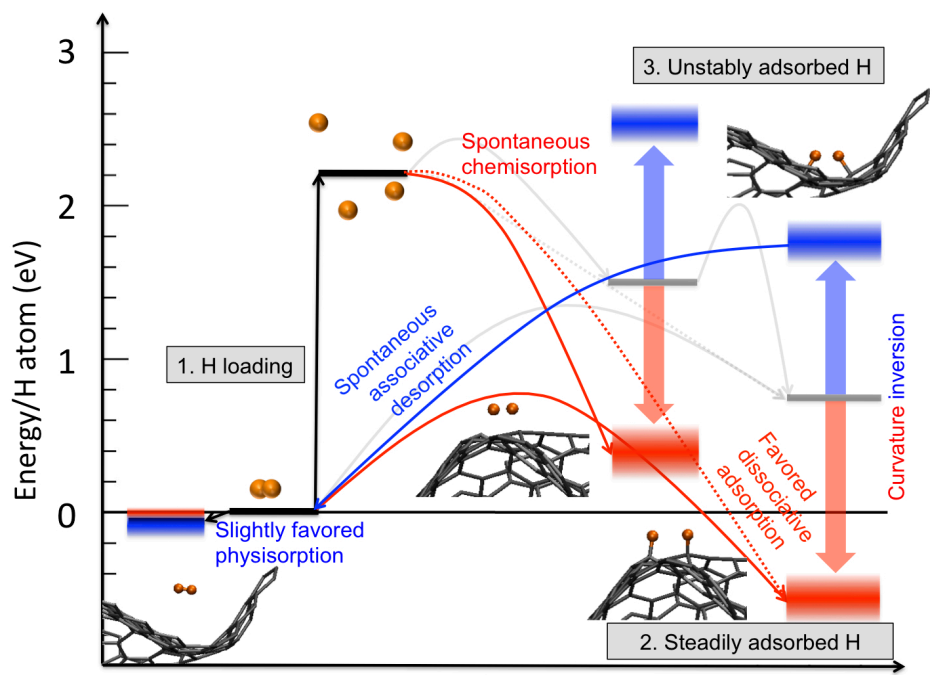

**Figure 4**
Change of the energy levels due to local graphene curvature (to compare with Figure 2). Convexities (red) stabilize the chemisorbed H and slightly destabilize the physisorbed $H_2$. Concavities (blue) act in the opposite way. Convexity causes the complete elimination of chemisorption barriers and the reduction of dissociative adsorption barrier, concavity eliminates the associative desorption barrier, and favours the physisorption at very low temperatures. Ball-sticks representations of bound and physisorbed hydrogen on curved graphene are also reported. Energy reference levels and scales as in Fig. 2.

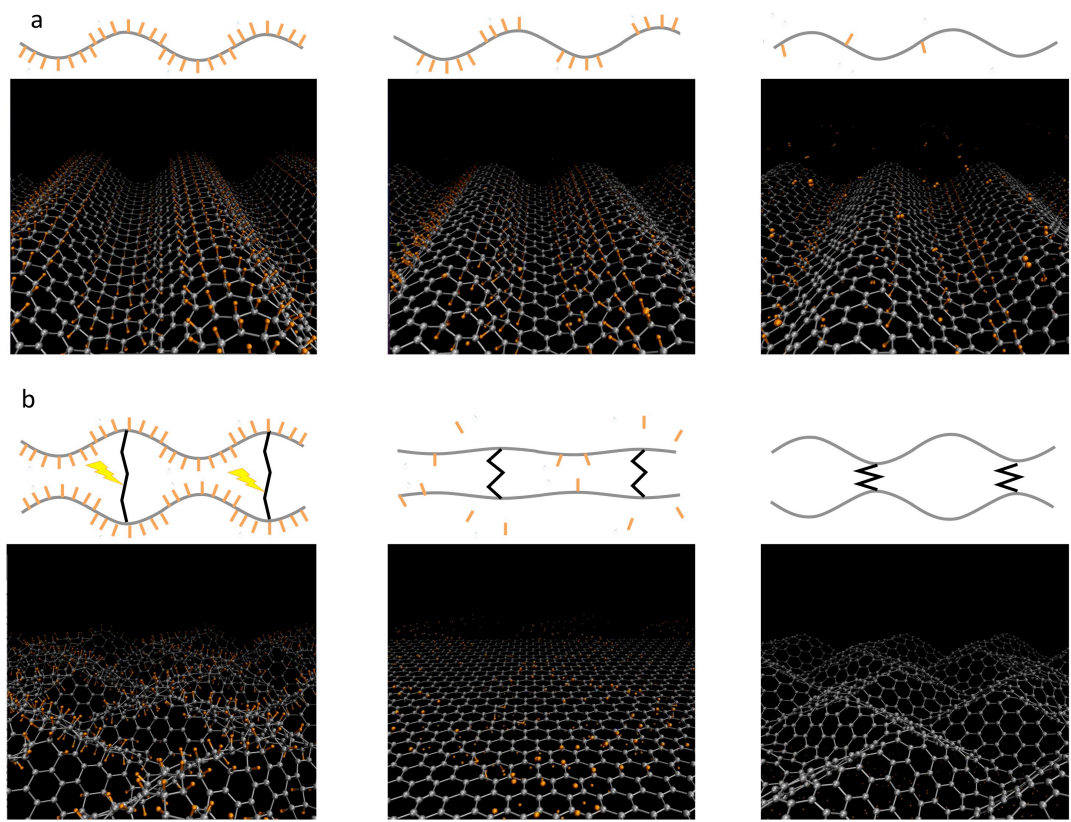

**Figure 5**
Possible protocols for hydrogen desorption. (a) Travelling wave: the curvature inversion is realized after half a period of a wave of mechanical distortion (an acoustic transverse phonon). This could be obtained using piezoelectric substrates. (b) the inversion is realized directly by changing the size of properly designed intercalants with external stimuli (e.g, optical excitation). The intercalants could function also as spacers.

**Notes and references**